*Article*

# Dynamical Embedding of Single-Channel Electroencephalogram for Artifact Subspace Reconstruction

Doli Hazarika [1], K. N. Vishnu [1], Ramdas Ransing [2] and Cota Navin Gupta [1,*]

1. Neural Engineering Lab, Department of Biosciences and Bioengineering, Indian Institute of Technology, Guwahati 781039, India; dhazarika@iitg.ac.in (D.H.); kvishnu@iitg.ac.in (K.N.V.)
2. Department of Psychiatry, Clinical Neurosciences, and Addiction Medicine, All India Institute of Medical Sciences, Guwahati 781101, India; ramdasransing@aiimsguwahati.ac.in
* Correspondence: cngupta@iitg.ac.in; Tel.: +91-361-258-2232

**Abstract:** This study introduces a novel framework to apply the artifact subspace reconstruction (ASR) algorithm on single-channel electroencephalogram (EEG) data. ASR is known for its ability to remove artifacts like eye-blinks and movement but traditionally relies on multiple channels. Embedded ASR (E-ASR) addresses this by incorporating a dynamical embedding approach. In this method, an embedded matrix is created from single-channel EEG data using delay vectors, followed by ASR application and reconstruction of the cleaned signal. Data from four subjects with eyes open were collected using Fp1 and Fp2 electrodes via the CameraEEG android app. The E-ASR algorithm was evaluated using metrics like relative root mean square error (RRMSE), correlation coefficient (CC), and average power ratio. The number of eye-blinks with and without the E-ASR approach was also estimated. E-ASR achieved an RRMSE of 43.87% and had a CC of 0.91 on semi-simulated data and effectively reduced artifacts in real EEG data, with eye-blink counts validated against ground truth video data. This framework shows potential for smartphone-based EEG applications in natural environments with minimal electrodes.

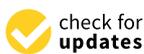

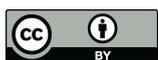

## 1. Introduction

Electroencephalography (EEG) is a non-invasive method employed for capturing the electrical patterns generated by cortical neurons, achieved by positioning electrodes on the scalp [1]. EEG amplifiers, known for their portability and capacity to offer precise temporal resolution in signal recording, establish EEG as the optimal brain imaging tool for assessing human brain activity during motion [2]. In recent years, there has been an increasing interest in conducting EEG experiments in natural environments using smartphones, marking a significant shift in EEG experimentation [3]. Smartphone-based EEG offers several advantages, including portability and affordability, positioning it as a promising next-generation technique for real-time brain activity investigation [4]. Furthermore, as technology continues to advance, these systems have evolved to feature low instrumentation and computational complexity [5,6]. Notably, portable EEG devices equipped with a single EEG channel have gained widespread use in non-laboratory and non-clinical applications, reflecting their practicality [7,8]. These devices have found utility in diverse domains, ranging from BCI research to driver fatigue detection and the study of various brain disorders [9–11].

However, EEG is susceptible to contamination by artifacts. Non-physiological artifacts can include high impedance, faulty electrodes, or noise from surrounding electrical equipment. Physiological and biological artifacts, including blinks, eye movements, muscular activity, and heart-related signals, pose a substantial challenge in EEG signal analysis, making their removal a primary focus when addressing EEG artifacts [12,13]. The activity of the eyes including blinks and saccades produce large amplitude changes in prefrontal (Fp1 and Fp2) electrodes. As we traverse from front to the back of the scalp, the eye-blink amplitude decreases. Typically, these artifacts exhibit an amplitude of 500 microvolts and a frequency below 20 Hz [12,13], characteristics that are also linked to upper-limb movements and drivers' cognitive states [14–16]. Electrical activity produced by muscle movements including jaw clenching, swallowing, and changes in facial expression results in large amplitude changes in EEG signal. Nonetheless, improper artifact filtering can impact the signal in terms of both its temporal and frequency characteristics, potentially leading to a loss of critical information, which could, in turn, jeopardize the effectiveness of various natural environment EEG application.

The artifact subspace reconstruction (ASR) algorithm is an adaptive spatial filtering method for removal of artifacts from EEG signals, developed and patented by C.A.E. Kothe and T.P. Jung in 2016 [17]. This method performs a Principal Component Analysis (PCA) on the EEG data using a sliding window approach. In the initial step, ASR automatically derives reference data from the raw signal based on the distribution of signal variance. Subsequently, it establishes thresholds for identifying artifact components by considering the standard deviation across the principal components of the windows, which is then multiplied by a user-defined parameter 'k'. Then, ASR identifies and eliminates artifact components within each time window if their principal component surpasses the rejection threshold. Finally, the method reconstructs the cleaned signals using the remaining data [18]. According to [19], the 'k' parameter dictates the aggressiveness of the faulty data removal process. A smaller 'k' results in a more aggressive removal procedure. An enhanced variant called Riemannian ASR employs manifold techniques for computing covariance matrices, a method proven to be effective for artifact removal [20]. In an investigation involving motor imagery EEG data [21], it was observed that an ASR technique with default settings performs better than the Independent Component Analysis (ICA) and PCA methods. In [18], researchers illustrated the efficacy of ASR as an automated approach for removing artifacts from EEG data collected during attention tasks in a driving simulator. Additionally, ref. [22] highlights the application of ASR to EEG data recorded during activities such as fast walking and maintaining a single-leg stance. Lately, ASR has been incorporated into the Smarting Pro smartphone application, enabling the automatic removal of artifacts from multiple channels [6]. However, existing ASR algorithms cannot be applied to single-channel EEG recordings and their performance can be impaired when the number of channels is small [18].

In a study [23], the feasibility of employing Singular Spectrum Analysis (SSA) to mitigate eye-blink artifacts in single-channel EEG data was explored. The technique is applied to extract low-frequency, oscillatory, and noisy components from singular time series data [24]. However, traditional SSA requires a crucial step in which the relevant signal eigenvectors must be identified. A novel set of criteria for the selection of these eigenvectors, crucial for reconstructing the desired signal, was also introduced [24]. SSA was subsequently integrated into the Adaptive Filter (AF) framework to enhance performance in [25]. Furthermore, single-channel EEG recordings have been subjected to ICA after undergoing SSA processing [26]. In a recent development, SSA was employed as a smoothing filter to mitigate the Electrooculogram (EOG) artifacts present in EEG signals [27]. A study [28] explored the integration of SSA-ICA and wavelet thresholding techniques to eliminate EOG artifacts in single-channel contaminated EEG signals. Reference [29] introduced a versatile approach for EEG artifact removal with limited supervision. They presented an innovative wavelet-based technique enabling the elimination of artifacts from single-channel EEG through a data-driven adjustment of wavelet coefficients. Their method demonstrates the ability to dynamically reduce artifacts of varying types. Nevertheless, the utilization of the aforementioned SSA algorithms into Android smartphone applications for artifact removal from EEG signals remains an area that has not been explored.

The majority of the approaches for eliminating eye-blink artifacts as mentioned earlier are primarily employed for offline artifact elimination. However, in situations like natural environment EEG experiments and epilepsy monitoring, where real-time signal processing is essential, it becomes imperative that artifact removal algorithms are capable of handling real-time processing. Consequently, to accommodate the needs of real-time artifact removal, these methods or algorithms must meet specific criteria. The foremost requirement is that the algorithm must operate automatically, without the need for any manual intervention. Secondly, it is crucial to use minimum number of electrodes for natural environment applications as this can cause discomfort and inconvenience to the subject during prolonged EEG recordings. One of the major advantages of using single-channel EEG is its simplicity and ease of use, as it requires less setup time and minimal equipment compared to multiple-channel EEG. Additionally, it offers cost-effective solutions for researchers, especially those working with limited resources, while still providing valuable data. For instance, studies have shown that single-channel EEG can be used to identify cognitive states such as driver drowsiness detection [30], as well as brain activity associated with specific mental disorders like depression and anxiety [31]. Lastly, for real-time implementation on a smartphone app, the artifact removal algorithm should have minimal computational complexity to ensure that it does not introduce unacceptable time delays [3].

The dynamical embedding concept to perform ICA using single-channel EEG was introduced in [32] for the separation of ocular artifacts. Pseudo-multichannel data called the embedding matrix was created using delayed vectors spanning a few seconds, from a single-channel EEG recording. This embedding matrix was used as the input to ICA. Embedding is also a fundamental part of SSA, allowing for the separation of underlying artifact components from single-channel EEG [24]. ASR has been successfully implemented on an android smartphone [6] for addressing real-time artifacts, leveraging multiple channel inputs. However, to the best of our knowledge, no work has studied ASR for single-channel EEG. Therefore, our primary objective was to investigate the effectiveness of embedding as a method for implementing ASR on single-channel EEG data. We achieved this by creating an *EmbeddedMatrix* from a single-channel EEG signal and then applying ASR. This framework may be a potential solution for artifact reduction on a smartphone for natural environment EEG experiments. As a first step, to assess the performance of our novel E-ASR framework, we focused on metrics (calculated before and after E-ASR framework) such as relative root mean square error, correlation coefficient, average power ratio, and reduction in number of eye-blinks. We evaluated the performance on semi-simulated and real EEG signals. Figure 1 illustrates the graphical abstract of the proposed framework.

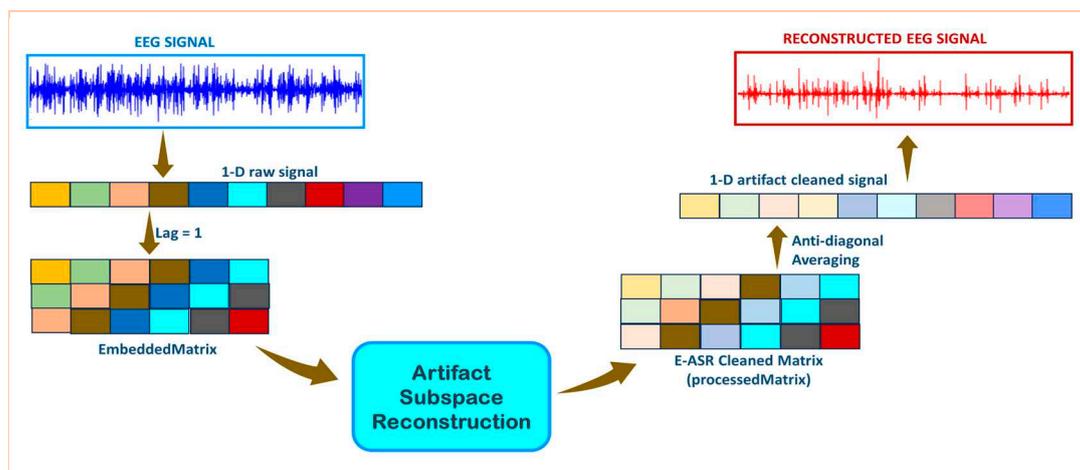

**Figure 1.** Graphical abstract for the proposed E-ASR framework on a single electroencephalogram channel. Each color represents a single time point.

## 2. Materials and Methods

*2.1. Data Acquisition and Pre-Processing*

The Indian Institute of Technology Guwahati Human Ethics Committee approved this research work. It was conducted in accordance with the principles embodied in the Declaration of Helsinki and in accordance with local statutory requirements. We obtained EEG data for four (two male and two female, with a mean age of 28 years and standard deviation of 4.33) subjects with the CameraEEG android application, which synchronously records video and EEG data [33]. The app is compatible with all Android smartphones running Android OS Lollipop or higher. We employed the CameraEEG app alongside an EasyCap 24-channel headcap [34] and the mBrainTrain Smarting device [5]. The subjects were asked to keep their eyes open for 5 min. The acquired video and EEG data were saved as mp4 and bdf files, respectively, on the smartphone's memory. The sampling frequency was set at 500 Hz. Figure 2 shows the photographs of the recording setup used for this work.

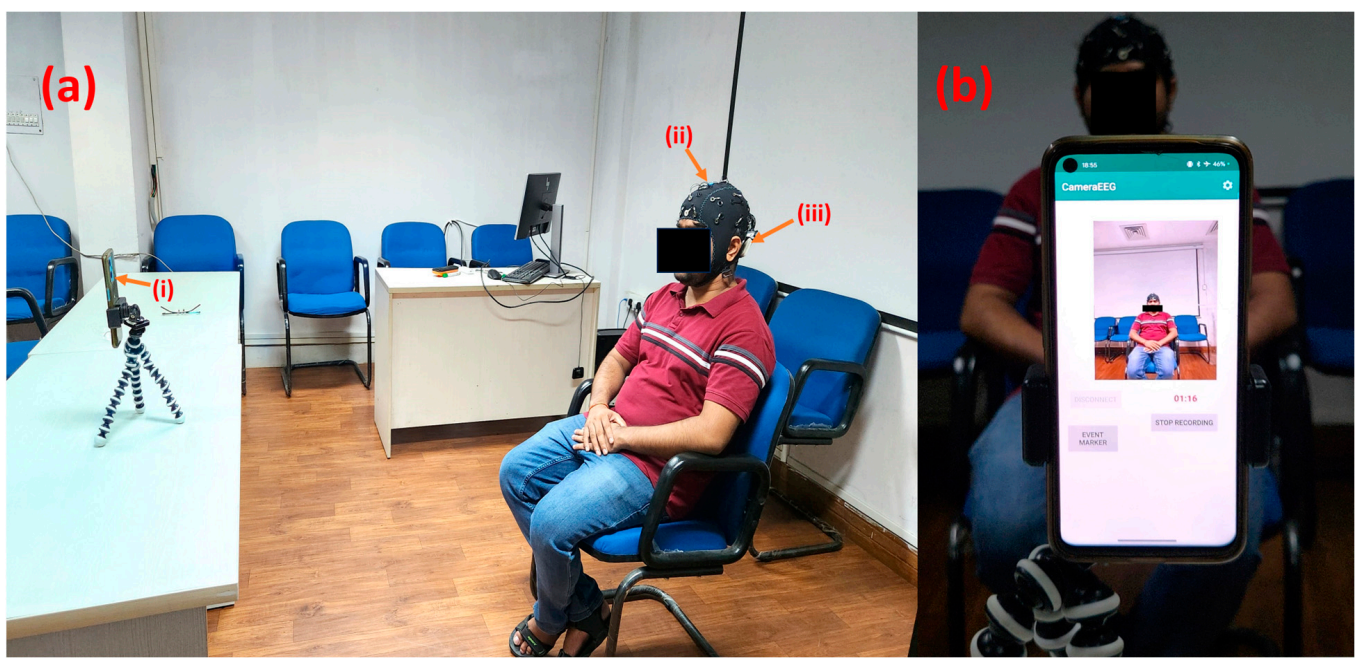

**Figure 2.** (**a**) The data recording setup for the resting-state eyes-open task, featuring (**i**) a OnePlus Nord CE 2 Lite 5G smartphone placed on a tripod in front of the subject, (**ii**) an EasyCap 24-channel EEG cap, and (**iii**) an mBrainTrain Smarting device mounted on the EEG cap. (**b**) The CameraEEG Android app running on the smartphone, recording synchronized EEG and video data.

The 5 min eyes-open data from the all the subjects is considered here throughout for the analysis. The Fp1 and Fp2 channels from the EEG data were selected and considered as single-channel EEG signals for further use, as they would contain the most eye-blinks and eye movement-related artifacts. We normalized the data using zero-centered normalization [18]. Following this, the single-channel signal was filtered by a band-pass filter (0.5–100 Hz) and a notch filter was used to remove 50 Hz line noise (electrical shifts) [18,35]. There was no linear trend observed in the signal through visual inspection; hence, detrending of the signal was not considered. This might be a potential limitation of the current study. The MATLAB codes used were developed using MATLAB version 2022b (MathWorks, Natick, MA, USA) on a system with an Intel® Core (TM) i7-8700 CPU @ 3.19 GHz and 16 GB memory.

*2.2. Construction of Multichannel EEG Matrix Using Embedding Approach*

The embedding matrix is a way of representing the temporal structure of an EEG signal [32]. It is created by making a series of delay vectors on a single channel EEG data. This matrix captures information about underlying EEG generators based on single-channel data [36,37]. Let us consider a single-channel electroencephalogram signal as $x = [x(1), x(2), \ldots, x(N)]$, where $N$ is the total number of samples. Then, the multidimensional series can be written as Equation (1):

$$X = \begin{bmatrix} x(1) & x(2) & \cdots & x(K) \\ x(2) & x(3) & \cdots & x(K+1) \\ \vdots & \vdots & \ddots & \vdots \\ x(M) & x(M+1) & \cdots & x(N) \end{bmatrix} \quad (1)$$

where $M$ is the embedding dimension and $K = N - M + 1$. If $f_s$ is the sampling frequency of the signal and $f_L$ is the lowest frequency of interest, then the embedding dimension $M$ [32] can be determined by Equation (2):

$$M \geq \frac{f_s}{f_L} \quad (2)$$

In our approach, the time lag is set to 1, which is supported by empirical evidence from [32]. We shall henceforth refer to Equation (1) as *EmbeddedMatrix*.

The embedding dimension ($M$) is a crucial parameter in decomposing a time series data. It determines the number of lagged components of the time series. Selecting an appropriate embedding dimension $M$ is essential because it influences the quality of the decomposition and the ability to extract meaningful information from the time series. If $M$ is too small, important information may be lost, leading to incomplete decomposition. On the other hand, if $M$ is too large, it can lead to overcomplexity and noise in the decomposition, making it harder to extract meaningful components [32]. Choosing the optimal $M$ often involves a balance between capturing important patterns and minimizing noise [32].

*2.3. Artifact Subspace Reconstruction*

In the first step of ASR (i.e., calibration phase), the EEG data ($X$) is input for the *asr_calibrate* function along with the sampling frequency in Hertz (Hz) to construct the calibration data and determine rejection thresholds from the calibration data [18]. To do so, the covariance matrix of $X$ is calculated. Mixing matrix ($M_C$) is calculated as the square root of covariance matrix as in Equation (3).

$$M_C M_C^T = Cov(X) \quad (3)$$

The eigenvalue decomposition of $M_C$ results in eigenvectors ($V_C$) and eigenvalues ($D_C$). The principal component space is calculated as Equation (4).

$$Y_C = X * V_C \quad (4)$$

Component-wise root mean square (RMS) values with a non-overlapping sliding window of 1 s [18] are calculated and transformed into z-score. ASR selects the windows with z-score in range of $-3.5 < z < 5.5$ [18], defines them as clean (artifact-free) sections, and concatenates them to generate the calibration data. From the clean sections of each principal component, the mean ($\mu$) and standard deviation ($\sigma$) of each component are calculated. The threshold of each component is calculated as Equation (5), where $i$ refers to the principal component number and $k$ is the cut-off parameter, whose value was 17 [18].

$$T_i = \mu_i + k\sigma_i \tag{5}$$

The threshold matrix $T$ is the matrix product of diagonal matrix of threshold values $T_i$ and the transpose of eigenvectors $V_C$. The threshold matrix $T$ and the mixing matrix $M_C$ are the outputs of this calibration phase, which are stored in a variable 'state' [18].

In the second step (i.e., process phase) of ASR, eigenvalue decomposition is performed on data within a sliding window (0.5 s) [18] to obtain the eigenvalues ($D_T$) and eigenvectors ($V_T$) of the data. Then *asr_process* function applies thresholds determined in the calibration phase to create $X_T$ [18]. If an eigenvalue within that sliding window exceeds the threshold, its corresponding eigenvector is removed. The leftover eigenvectors ($V_{trunc}$) are hence truncated and the data are reconstructed within that window according to Equation (6) from [18].

$$(X_T)_{clean} = M_C \left( V_T^T M_c \right)^+_{trunc} V_T^T X_T \tag{6}$$

Due to the rank reduction from truncation, the Moore–Penrose pseudoinverse (+) ensures a reliable reconstruction by finding an optimal solution that minimizes reconstruction error. This approach allows for accurate data recovery, even when the matrix is singular or lacks full rank. The clean windows (($X_T)_{clean}$) are concatenated to form the clean artifact-free signal $X_{Clean}$. More details on the ASR algorithm can be found in [17,35].

*2.4. Proposed Method: Embedded Artifact Subspace Reconstruction*

The embedded matrix is created by time lagging the pre-processed single-channel EEG data. We determined the embedding dimension ($M$) to be large enough to capture the information content in the signal. For the EEG signals described in this study, we derived $M$ using the equation given in (2). The pre-processed 1D EEG data were transformed into *EmbeddedMatrix* as explained in Section 2.2, with time lag as 1. ASR is applied on the *EmbeddedMatrix* using the MATLAB codes available in EEGLAB [38] as an open-source plug-in function *clean_rawdata*. The output of application of ASR on *EmbeddedMatrix* is the *processedMatrix* of the same dimensions. Anti-diagonal averaging [37] is then applied to *processedMatrix* to reconstruct the E-ASR-cleaned signal.

## 3. Performance Metrics

We evaluated the effectiveness of our method on a semi-simulated single-channel EEG signal using two well-known metrics: the relative root mean square error (RRMSE) and the correlation coefficient (CC). The RRMSE is a commonly used measure for assessing the performance of artifact removal techniques in semi-simulated EEG data [39]. The correlation coefficient (CC), a statistically based metric, indicates the relationship between two signals and is employed to evaluate the effectiveness of the artifact removal process [39]. A higher CC value suggests a stronger linear relationship, implying better performance of the artifact removal method [39]. These metrics are calculated between the ground truth signal, which is free of eye-blink artifacts, and the artifact-cleaned signal [39,40]. Additionally, we analyzed the average power ratio across different frequency bands and the reduction in eye-blinks [40]. This analysis involves dividing the average power of each frequency band by the overall average power of the entire signal to gauge the relative contribution of each frequency band to the total signal strength [40]. In addition to these parameters, we employed blink count estimation using an amplitude threshold technique, as detailed below.

*Blink Count Estimation Using Amplitude Threshold*

The eye-blinks in the signal are calculated using amplitude, which ensures that any large-amplitude artifacts arising from the single channel in prefrontal region can be attributed to eye-blinks or eye movements. Any signal amplitude value exceeding a threshold would be considered as an eye-blink [41]. We experimented by varying the constant parameter from 1 to 10 for Fp1 and Fp2 electrodes of three subjects. The ground truth of eye-blink count obtained from the CameraEEG [33] video data was used for cross-examining the various counts from the amplitude threshold formula. The counted eye-blinks were also manually cross-checked with the CameraEEG video data. We observed that the constant parameter of 6 matched with the ground truth eye-blink count. A minimum distance between values exceeding this blink amplitude threshold was used to differentiate one eye-blink from another and to avoid multiple detections of a single blink event [41]. The minimum peak distance and threshold values were varied until the expected separation and occurrence of eye-blinks was reached for the EEG signals used in this study. Therefore, the blink amplitude threshold used was determined as given in Equation (7) and the minimum distance between subsequent peaks was set at 250 ms [42].

$$Blink\ Amplitude\ Threshold = 6 \times \frac{\sum_{i=1}^{n}|x_i|}{n} \tag{7}$$

where $x$ is the amplitude of the signal at sample number $i$ with total sample points $n$.

The eye-blinks are counted in this manner for the single channel EEG data before and after application of E-ASR. The large amplitude artifacts were calculated for each subject (Fp1 and Fp2). The percentage reduction in artifacts was calculated as given in Equation (8).

$$Percentage\ reduction = \frac{Before\ ASR - After\ ASR}{Before\ ASR} \times 100 \tag{8}$$

## 4. Results

### 4.1. Construction of Semi-Simulated Single Channel EEG and Eye-Blink Artifact

We created a semi-simulated dataset as given in [27] for testing the proposed method. The EEG signals were acquired using the Smarting device, sampled at 500 Hz with a resolution of 24 bits [5]. Two clean EEG segments about ten seconds long, without eye-blinks, are manually identified and extracted from Fp1 channel of subject 4. These two segments are then replicated and concatenated to a form 1-minute-long ground truth single-channel EEG signal. Further, two eye-blink artifacts are manually segmented and extracted from the same dataset, which is about 2 s long. To achieve a consistent signal length of 10 s, we extended the isolated eye-blink segments by adding zeros on both ends. Each clean segment is combined with one eye-blink segment using Equation (9):

$$z = s + \propto m \tag{9}$$

where $z$ is the semi-simulated contaminated EEG segment, $s$ is the clean EEG segment, $m$ is the eye-blink segment, and $\propto$ is the mixing coefficient that controls the signal-to-noise ratio (SNR) of the constructed noisy signal [39,40]. The mixing coefficient $\propto$ can be calculated using Equation (10):

$$SNR = 10 log\left(\frac{RMS(s)}{RMS(\propto m)}\right) \tag{10}$$

The SNR values were considered within the range −7 dB to 2 dB to calculate $\propto$ [40]. The root mean square (RMS) is given by Equation (11):

$$RMS(s) = \sqrt{\frac{1}{n}\sum_{i=1}^{n} s_i^2} \tag{11}$$

Four variations were created by combining these clean and eye-blink segments in different orders. Finally, these were combined to create a 1 min semi-simulated contaminated EEG signal. Figure 3 shows the schematic representation of the steps involved in creating semi-simulated signal.

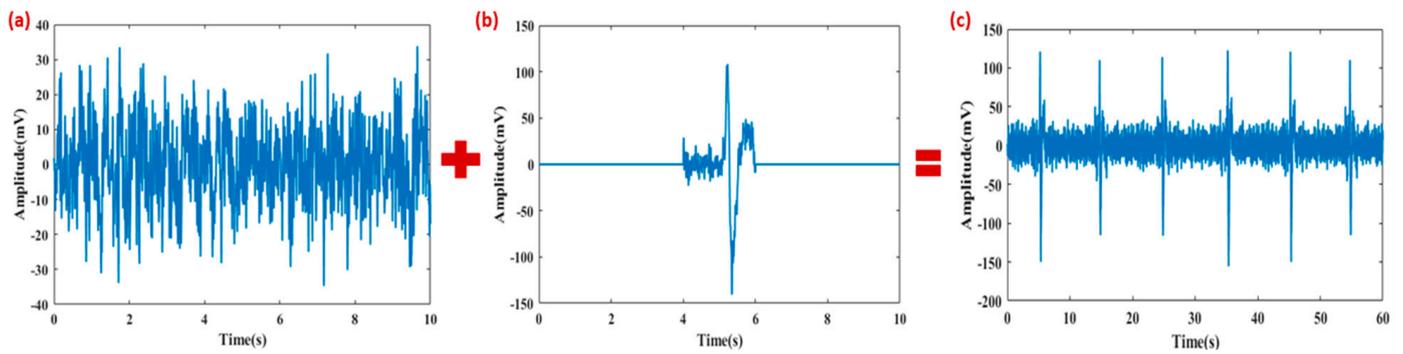

**Figure 3.** Framework for creating semi-simulated signal: (**a**) 10 s clean EEG segment from subject 4, (**b**) eye-blink, and (**c**) superposition of both clean EEG and eye-blink to create semi-simulated signal.

*4.2. Results with Semi-Simulated Single Channel EEG Signal*

The superposition plots of semi-simulated contaminated EEG, E-ASR-cleaned, and ground truth signal using the proposed method are shown in Figure 4.

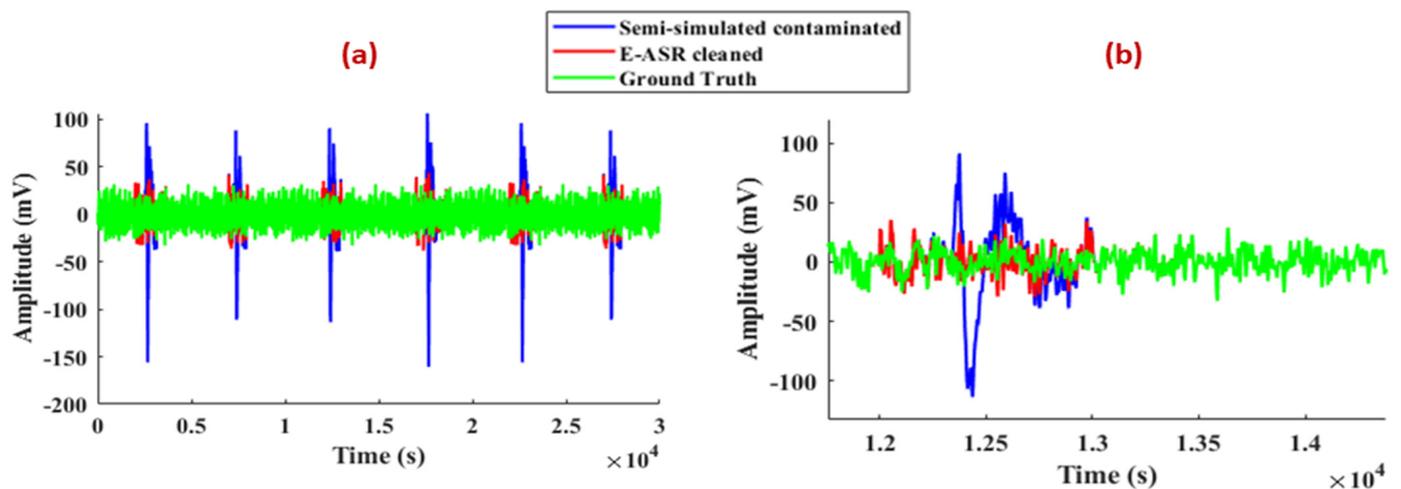

**Figure 4.** The superposition plots of semi-simulated contaminated EEG, E-ASR-cleaned, and ground truth signal using the proposed algorithm: (**a**) plot for 1 min time duration signal; (**b**) zoomed version of (**a**) showing one eye-blink.

It can be observed from Figure 4b that the eye-blinks are visibly reduced after applying E-ASR to the contaminated signal. An RRMSE of 43.87% and a CC of 0.91 were achieved for E-ASR when applied to the semi-simulated signal. To enable a comparison with the state-of-the-art ASR algorithm, an additional semi-simulated signal was generated, forming a two-channel dataset. ASR was then applied to this two-channel semi-simulated dataset, and its performance was assessed by calculating the RRMSE and CC for the first channel of the ASR-cleaned data, yielding an RRMSE of 56.82% and a CC of 0.85. To assess how well our method performed across different EEG frequencies, we analyzed the average power distribution within each band relative to the entire spectrum for eyeblink artifact removal (Table 1). We focused on the delta (0.5–4 Hz), theta (4–8 Hz), alpha (8–13 Hz), beta (13–30 Hz), and gamma (30–100 Hz) bands, encompassing the whole 0.5–100 Hz range. We also counted the eye-blinks using Equation (7); the semi-simulated contaminated signal contained six eye-blinks and E-ASR effectively removed all of them.

**Table 1.** Average power distribution of EEG frequency bands for semi-simulated contaminated signal, ground truth, and E-ASR-cleaned signal.

| Frequency Bands | Contaminated EEG | E-ASR-Cleaned EEG | Ground Truth EEG |
|---|---|---|---|
| Delta (0.5–4) | 0.63 | **0.21** | 0.23 |
| Theta (4–8) | 0.14 | **0.14** | 0.15 |
| Alpha (8–13) | 0.04 | **0.10** | 0.10 |
| Beta (13–30) | 0.15 | **0.44** | 0.41 |
| Gamma (30–100) | 0.04 | **0.12** | 0.10 |

*4.3. Results with Real EEG Signals*

Unlike simulated data where we have a perfect version of the signal, real EEG recordings lack a ground truth. Therefore, to assess our method's performance, we manually identified sections of the recordings (from 5 min eyes-open data) that did not contain artifacts and concatenated them to obtain an artifact-free signal of 1 min. To evaluate the effectiveness of our method, we calculated the RRMSE and CC between the artifact-free signal and its E-ASR-cleaned version, as shown in Table 2. It demonstrates that the method consistently achieves high correlation (CC values close to 0.9) while maintaining reasonable error levels (RRMSE) across all subjects and channels. We also show the comparison of average power ratio between these two signals for all subjects in Figure 5.

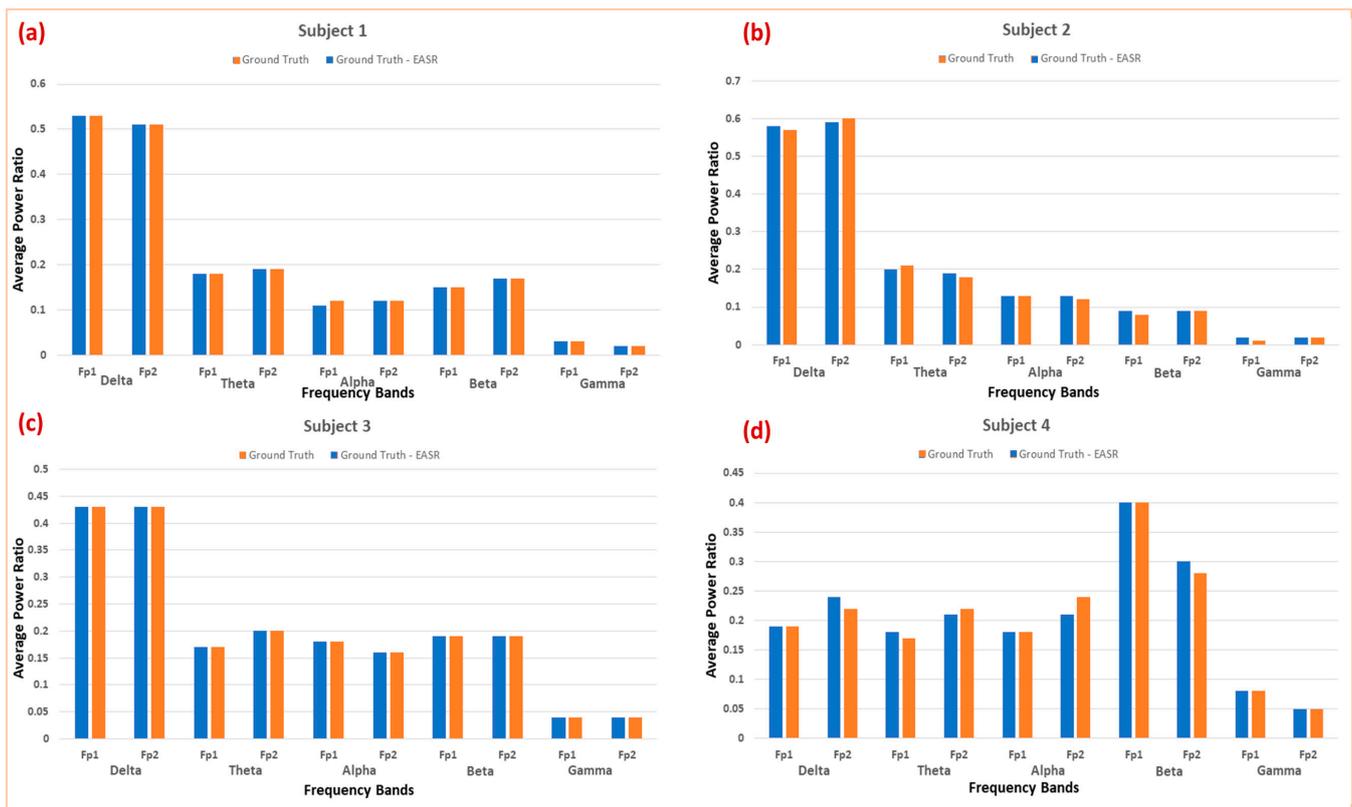

**Figure 5.** Comparison of average power ratio between the artifact-free signal and its E-ASR-cleaned version for all subjects. The E-ASR algorithm successfully restored the power distribution across the EEG spectrum for each subject (**a**–**d**).

We sought to evaluate the performance of the proposed EASR algorithm against the original ASR algorithm. The 24-channel data we originally collected was cleaned by the ASR algorithm. In contrast, for application of E-ASR, a single channel was used from the 24-channel montage. The ASR-cleaned Fp1 and Fp2 channels were considered for the time domain comparison with the proposed EASR algorithm in Figure 6. Eye-blinks exhibit a distinctive peak that becomes evident in time-domain EEG signals. Notably, the distinct peaks corresponding to eye-blinks in the signal are eradicated after the application of E-ASR. Table 3 presents the number of eye-blinks recorded before and after the application of E-ASR for each subject. All subjects demonstrated a complete elimination of eye-blinks post-E-ASR, indicating a 100% reduction. The computational time reflects the duration taken for processing, varying slightly among subjects. Also, the generality of the framework is shown by considering different sampling frequency and electrode locations (Appendix A: Tables A1 and A2).

**Table 2.** RRMSE and CC between the artifact-free signal and its E-ASR-cleaned version.

| Subject | Channel | RRMSE (%) | CC |
|---|---|---|---|
| 1 | Fp1 | 41.06 | 0.91 |
|   | Fp2 | 38.72 | 0.92 |
| 2 | Fp1 | 47.37 | 0.88 |
|   | Fp2 | 52.07 | 0.86 |
| 3 | Fp1 | 46.26 | 0.89 |
|   | Fp2 | 45.98 | 0.89 |
| 4 | Fp1 | 43.53 | 0.90 |
|   | Fp2 | 40.45 | 0.91 |

**Table 3.** Change in number of eye-blinks before and after E-ASR on 1 min single-channel real EEG data for all subjects. Their computational time is also reported.

| Subject | Channel | No. of Eye-Blinks Before E-ASR | No. of Eye-Blinks After E-ASR | Percentage Reduction of Eye-Blinks (%) | Computational Time (Seconds) |
|---|---|---|---|---|---|
| Subject 1 | Fp1 | 9 | 0 | 100 | 5.2 |
|   | Fp2 | 9 | 0 | 100 | 4.9 |
| Subject 2 | Fp1 | 8 | 0 | 100 | 5.3 |
|   | Fp2 | 4 | 0 | 100 | 5.3 |
| Subject 3 | Fp1 | 14 | 0 | 100 | 5.6 |
|   | Fp2 | 16 | 0 | 100 | 5.4 |
| Subject 4 | Fp1 | 7 | 0 | 100 | 4.7 |
|   | Fp2 | 7 | 0 | 100 | 4.5 |

To illustrate the impact of applying E-ASR to a single channel, topographic plots were generated at a specific time point during which the subject exhibited an eye-blink. Figure 7A illustrates that in the absence of ASR application, distinct high-amplitude peaks (dark red regions) were observed in the prefrontal region from eye-blinks [43]. Upon applying single-channel E-ASR to Fp1, the resulting ASR-cleaned channel was used in the 24-channel EEG configuration for the purpose of generating topographic plots.

The associated scalp map in Figure 7B demonstrates the successful removal of the blink artifact from Fp1. However, Fp2 continued to have blink-related activity. E-ASR was also independently applied on both Fp1 and Fp2 electrodes, and we observed effective elimination of the eye-blink activity, as depicted in Figure 7C. This suggests that the single-channel E-ASR framework is reasonably effective in removing artifact content.

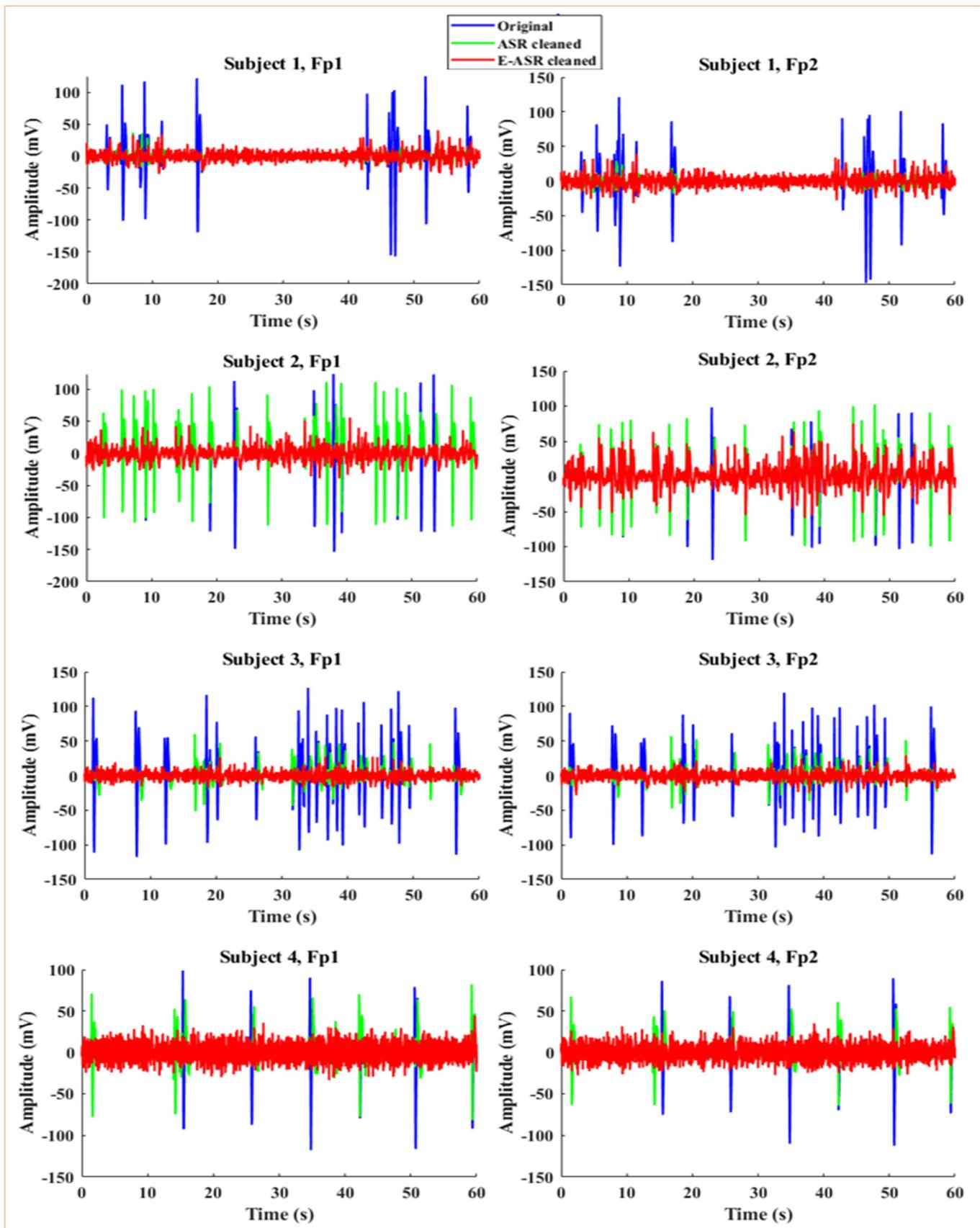

**Figure 6.** Time-domain comparison of original (blue), ASR-cleaned (green), and E-ASR-cleaned (red) on Fp1 and Fp2 channels across all subjects.

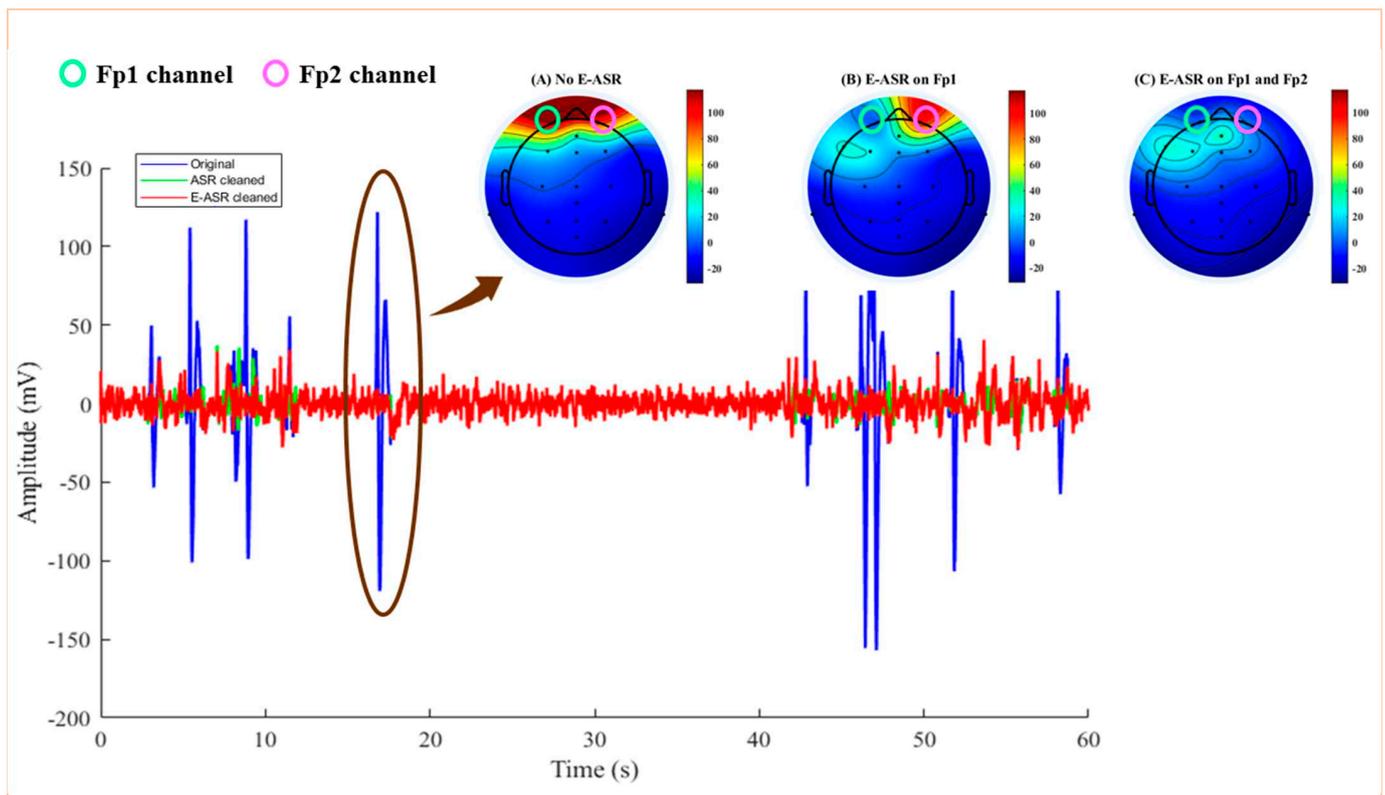

**Figure 7.** Spatial distribution of source activities at an eye-blink time point for (**A**) no E-ASR, (**B**) E-ASR only applied on Fp1, and (**C**) E-ASR applied on Fp1 and Fp2 channels of subject 1. Red indicates the presence of an eye-blink artifact whereas blue indicates the absence. The green circle indicates the location of Fp1 electrode and pink indicates the location of Fp2 electrode.

## 5. Discussion

The developed framework aims to explore the efficacy of a novel Embedded Artifact Subspace Reconstruction (E-ASR) for addressing artifact removal for single-channel EEG data. The concept draws inspiration from dynamical embedding, initially proposed for single-channel ICA in the separation of ocular artifacts [36]. This idea was extended to create an embedding matrix from single-channel EEG data for implementing an artifact subspace reconstruction algorithm. Notably, while ASR has been successfully applied in a multichannel setting on an android smartphone [6], this study investigates the implementation of ASR specifically for single-channel EEG data. The primary goal was to assess the performance of the E-ASR framework by employing metrics such as RRMSE, CC, average power ratio, and percentage reduction in eye-blinks. We used an embedding dimension of 90 and lag ($L = 1$) for the proposed work, as in [32,44].

The application of the E-ASR algorithm on semi-simulated EEG data demonstrated highly promising results (Figure 4). The algorithm successfully removed 100% of the eye-blink artifacts, as evidenced by the achieved RRMSE of 43.87% and a high correlation coefficient (CC) of 0.91. These metrics suggest that the E-ASR method not only effectively eliminates artifacts but also retains the essential features of the original EEG signal, which is crucial for maintaining data integrity. To compare its performance with the state-of-the-art ASR algorithm, a two-channel semi-simulated dataset was generated. ASR was applied to this dataset, and its performance was evaluated by calculating the RRMSE and CC for the first channel of the ASR-cleaned data. This resulted in a higher RRMSE of 56.82% and a lower CC of 0.85. These findings suggest that E-ASR outperforms ASR, offering a more optimal balance between artifact reduction and signal preservation. Eye-blink artifacts predominantly interfere with the low-frequency EEG bands (0–12 Hz) [27,42], often leading to a shift in power distribution towards the delta band and subsequently weakening other frequency bands. This effect was evident when we introduced eye-blink artifacts into the semi-simulated data. However, the E-ASR algorithm managed to restore the power balance across the EEG spectrum, as observed in Table 1. Specifically, the power distribution in the delta, theta, alpha, beta, and gamma bands of the E-ASR-cleaned signal closely matched that of the ground truth, indicating that our method effectively mitigates the distortions caused by eye-blinks without compromising the inherent frequency characteristics of the EEG signal.

Real EEG recordings inherently lack a definitive ground truth, making the evaluation of artifact removal methods particularly challenging. In this study, we addressed this issue by manually constructing a 1 min artifact-free signal from clean sections of the EEG recordings. By comparing this signal with the E-ASR-cleaned version, we calculated the relative root mean square error (RRMSE) and correlation coefficient (CC) to quantify the performance of our method. The obtained mean RRMSE of 44.43% and a CC of 0.89 (from Table 2) indicate that our E-ASR algorithm performs consistently well in preserving the underlying signal characteristics while effectively reducing artifacts.

In real EEG data, the variability and complexity of eye-blinks are more pronounced, making artifact removal more challenging. Despite this, the E-ASR algorithm demonstrated robustness when applied to real EEG signals (Figure 6), effectively eliminating the distinct peaks associated with eye-blinks. This result suggests that the E-ASR method is capable of handling the dynamic nature of eye-blink artifacts in real-world scenarios, further emphasizing its practical applicability. The complete elimination of eye-blink artifacts, as demonstrated by the 100% reduction in eye-blinks across all subjects (Table 3), underscores the efficacy of the E-ASR algorithm. Eye-blinks are particularly problematic in EEG analysis, as they can obscure the brain activity of interest. Our results not only confirm the robustness of E-ASR in mitigating such artifacts but also highlight its potential utility in improving the quality of EEG data for subsequent analyses, such as event-related potentials or brain connectivity studies.

The results from applying E-ASR to 1 min segments of real EEG data (Table 3) revealed a complete (100%) reduction in eye-blinks. Although we have not yet conducted real-time implementation, we have provided the computational time required for our algorithm to run on a desktop computer using MATLAB software, version 2022b. The average processing time was measured at 5.14 s for 1 min of EEG data. While the original ASR algorithm requires slightly less computational time for the same data with equivalent sampling frequency, it is important to note that ASR cannot be applied to single-channel data. The difference in processing time primarily arises from the embedding and reconstruction steps in the E-ASR algorithm. Overall, the balance between effectiveness and processing time underscores the practicality of the E-ASR algorithm in mitigating eye-blink artifacts without significantly compromising information.

Additionally, the comparison of the E-ASR performance with the traditional ASR algorithm provided valuable insights into the advantages of our method. While ASR demonstrated some effectiveness in cleaning the EEG data, the persistence of eye-blink activity in Figure 6 after its application highlights the limitations of conventional methods. In contrast, the E-ASR algorithm's ability to fully eliminate eye-blinks from both Fp1 and Fp2 channels illustrates its superior capacity for artifact removal. The topographic plots (in Figure 7) further illustrate the impact of E-ASR in enhancing the clarity of EEG data. The absence of distinct high-amplitude peaks in the prefrontal region after applying E-ASR not only validates our approach but also emphasizes its potential relevance in clinical and research settings. The preservation of brain activity during periods of blinks could significantly improve the interpretability of EEG results and contribute to more accurate clinical assessments and cognitive neuroscience research.

## 6. Conclusions

In this paper, we present a novel approach for implementing ASR on single-channel EEG data. We generated a multichannel dataset by time-lagging prefrontal single-channel EEG data, known as dynamical embedding. We evaluated the effectiveness of the E-ASR method in removing eye-blink artifacts from this *EmbeddedMatrix*. Our findings reveal that the proposed E-ASR method achieved an average reduction of 100% in detected eye-blinks for the real dataset. This significant result underscores how eye-blink artifacts can interfere with the analysis of neural activity, potentially leading to misleading interpretations. The complete removal of these artifacts enhances the quality and reliability of EEG signals, making the data more suitable for subsequent analyses, such as cognitive assessments and clinical diagnostics. Furthermore, this achievement demonstrates the E-ASR algorithm's robustness in handling the variability inherent in real-world data, suggesting its potential as a standard pre-processing tool in EEG studies. Importantly, the algorithm also maintained the integrity of the underlying neural signals, as evidenced by consistent correlation coefficients and reduced relative root mean square error. Additionally, we used an embedding dimension value of 90 for the current dataset. Utilizing the ASR algorithm with a cut-off parameter of 17 ensured the preservation of brain activity.

The embedding dimension ($M$) plays a crucial role in the E-ASR algorithm. It determines the lowest frequency that can be extracted from the spectral decomposition of the *EmbeddedMatrix*. Exploring the effect of $M$ on performance metrics is a promising avenue for future research. Additionally, computational efficiency is also linked to the embedding dimension. As a result, reducing $M$ can potentially lead to faster processing times.

The framework's minimal channel requirements facilitate straightforward implementation, providing a practical advantage. Along with its performance and minimal electrode requirement, the novel single-channel E-ASR algorithm may be well suited for integration into a smartphone android application. We speculate that forthcoming natural environment EEG applications may see advantages in using this framework. To further validate these findings, future research should encompass more extensive investigations involving larger datasets.

**Author Contributions:** C.N.G. proposed the E-ASR algorithm idea to D.H. Then, the entire framework was built by D.H. and the verification was done by D.H., K.N.V. and C.N.G. The manuscript was written by D.H., C.N.G. and K.N.V.; R.R. contributed to discussions and revisions of the first written draft and later versions. All authors have read and agreed to the published version of the manuscript.

**Funding:** DH and KNV are funded by the MoE doctoral scholarship and NEWGEN-IEDC, DST, Project no.—D4, 2023–2024 from the Government of India. CNG's time was funded by NEWGEN-IEDC, Project no.—D4, 2023–2024. R.R's time was funded by Project ID: NHRP-Suicide/DL/MH-3/2024. IIT Guwahati and AIIMS Guwahati contributed equally to the APC funding.

**Institutional Review Board Statement:** This study was conducted in accordance with the Declaration of Helsinki and approved by the Institutional Ethics Committee of IIT Guwahati for recording data from human participants.

**Informed Consent Statement:** Informed consent was obtained from all subjects involved in this study.

**Data Availability Statement:** The MATLAB code developed for this work is available at our GitHub link below (accessed on 28 June 2024): https://github.com/NeuralLabIITGuwahati/E-ASR. Real electroencephalogram data for one subject is also provided as a .mat file.

**Conflicts of Interest:** The authors declare no conflicts of interest.

## Appendix A

**Table A1.** Eye-blinks and computational time corresponding to occipital, temporal, parietal, and midline central electrodes for all subjects.

| Subject | Channel | No. of Eye-Blinks Before E-ASR | No. of Eye-Blinks After E-ASR | Computational Time |
|---|---|---|---|---|
| Subject 1 | O2 | 0 | 0 | 4.88 |
|  | T8 | 0 | 0 | 4.82 |
|  | P7 | 0 | 0 | 4.98 |
|  | Cz | 1 | 1 | 4.90 |

**Table A1.** *Cont.*

| Subject | Channel | No. of Eye-Blinks Before E-ASR | No. of Eye-Blinks After E-ASR | Computational Time |
|---|---|---|---|---|
| Subject 2 | O2 | 2 | 2 | 4.67 |
|  | T8 | 0 | 0 | 4.65 |
|  | P7 | 1 | 1 | 4.69 |
|  | Cz | 0 | 0 | 4.73 |
| Subject 3 | O2 | 0 | 0 | 4.80 |
|  | T8 | 0 | 0 | 4.65 |
|  | P7 | 0 | 0 | 4.67 |
|  | Cz | 0 | 0 | 4.84 |
| Subject 4 | O2 | 0 | 0 | 4.64 |
|  | T8 | 0 | 0 | 4.62 |
|  | P7 | 0 | 0 | 4.72 |
|  | Cz | 0 | 0 | 4.67 |

**Table A2.** E-ASR results of 250 Hz sampling frequency for all subjects.

| Subject | Channel | No. of Eye-Blinks Before E-ASR | No. of Eye-Blinks After E-ASR | Computational Time (Seconds) |
|---|---|---|---|---|
| 1 | Fp1 | 6 | 0 | 1.44 |
|  | Fp2 | 9 | 0 | 1.45 |
| 2 | Fp1 | 8 | 0 | 1.29 |
|  | Fp2 | 5 | 2 | 1.50 |
| 3 | Fp1 | 15 | 0 | 1.42 |
|  | Fp2 | 16 | 0 | 1.32 |
| 4 | Fp1 | 7 | 0 | 1.35 |
|  | Fp2 | 7 | 0 | 1.23 |